# Multi-Scale, Multi-Resolution Brain Cancer Modeling

**Le Zhang, L. Leon Chen and Thomas S. Deisboeck \***

Complex Biosystems Modeling Laboratory, Harvard-MIT (HST) Athinoula A. Martinos Center for Biomedical Imaging, Massachusetts General Hospital, Charlestown, MA 02129, USA.

**\*Corresponding Author:**

Thomas S. Deisboeck, M.D.
Complex Biosystems Modeling Laboratory
Harvard-MIT (HST) Athinoula A. Martinos Center for Biomedical Imaging
Massachusetts General Hospital-East, 2301
Bldg. 149, 13th Street
Charlestown, MA 02129
Tel: 617-724-1845
Fax: 617-726-7422
Email: deisboec@helix.mgh.harvard.edu

# ABSTRACT

In advancing discrete-based computational cancer models towards clinical applications, one faces the dilemma of how to deal with an ever growing amount of biomedical data that ought to be incorporated eventually in one form or another. Model *scalability* becomes of paramount interest. In an effort to start addressing this critical issue, here, we present a novel *multi-scale* and *multi-resolution agent-based in silico* glioma model. While 'multi-scale' refers to employing an epidermal growth factor receptor (EGFR)-driven molecular network to process cellular phenotypic decisions within the micro-macroscopic environment, 'multi-resolution' is achieved through algorithms that classify cells to either active or inactive spatial clusters, which determine the resolution they are simulated at. The aim is to assign computational resources where and when they matter most for maintaining or improving the predictive power of the algorithm, onto specific tumor areas and at particular times. Using a previously described 2D brain tumor model, we have developed four different computational methods for achieving the multi-resolution scheme, three of which are designed to dynamically train on the high-resolution simulation that serves as control. To quantify the algorithms' performance, we rank them by weighing the distinct computational time savings of the simulation runs versus the methods' ability to accurately reproduce the high-resolution results of the control. Finally, to demonstrate the flexibility of the underlying concept, we show the added value of combining the two highest-ranked methods. The main finding of this work is that by pursuing a multi-resolution approach, one *can* reduce the computation time of a discrete-based model substantially while still *maintaining* a comparably high predictive power. This hints at even more computational savings in the more realistic 3D setting over time, and thus appears to outline a possible path to achieve scalability for the all-important clinical translation.

# 1. INTRODUCTION

In recent years, cancer modeling has become a rather popular interdisciplinary research topic in computational biology. Various approaches have been employed to move towards predictive oncology. Currently, such computational approaches include *continuum* [1-6], *discrete* [7-11] and *hybrid* models [12-19]. Although 'continuum' techniques can describe for example the change of cancer cell density [3, 4, 20], the diffusion of chemoattractant [16], heat transfer in hyperthermia treatment for skin cancer [21, 22], cell adhesion, and the molecular network of a cancer cell [23, 24] as an entire entity using differential equations, their ability to investigate single-cell behaviors and cell-cell interactions are very limited. Conversely, 'discrete' modeling can simulate (for example, via cellular automata [25]) the behavior of individual cancer cells as well as cell-cell and cell-extracelluar matrix (ECM) interactions [7, 10], but it fails when the inclusion and investigation of most fluid-physical aspects relevant to cancer becomes necessary [8]. To overcome the shortcomings of both modeling techniques, we employ a particular type of 'hybrid' approach, an *agent-based* model. The agent-based approach simulates *multi-scale* glioma growth and expansion [12-16, 26], thereby describing cancer as a complex dynamic, adaptive, and self-organizing system [27]. The advantage of such a model is that each cell is simulated as an agent equipped with an intracellular molecular signaling network that determines its phenotype on the microscopic scale [12, 13, 16]. This allows the investigation of the interactions among these cells, the interactions between the cells and ECM, as well as the impact of each cell's intracellular signaling dynamics on its spatio-temporal behavior within the micro-macroscopic environment. However, multi-scale agent-based modeling requires significant computational resources [28], especially when simulating millions of cancer cells within a realistic microenvironment. Our previous works [12-17, 29] therefore had to reduce the number of cancer cells used in the simulation as well as the ECM volume, thereby slowing the translation of this simulation platform into clinical applications. To temporarily resolve this *scalability* problem, one may choose to employ parallel computation [30] in an effort to reduce the computational workload for the continuum module. Nonetheless, it will be a critical prerequisite for this hybrid platform's clinical applicability to be efficient in terms of computational workload, while still maintaining sufficient predictive power. As a first step, here we develop four computational methods to classify cancer cells into either active or inactive clusters. In an active cluster, each cell's molecular profile is monitored ('high-resolution') to determine whether any phenotype switching is being triggered. In an inactive cluster, all cells are considered as an entity without tracking each individual cell's molecular profile ('low-resolution'). Next, we analyze and compare the computation time and predictive power of each method, determine their performance vis-à-vis the high-resolution control, and rank them accordingly. Finally, we introduce a combination method (consisting of the two computing methods with the highest separate ranking values) to demonstrate the flexibility of this multi-resolution approach for future extensions.

## 2. METHODS & TECHNIQUES

To start addressing the issue of scalability in agent-based cancer modeling algorithms [28], we present here several distinctively different computational methods designed to reduce the workload by utilizing selective spatial resolution, while simultaneously aiming to maintain sufficient overall predictive power. In the following sections, we will describe the modeling platform first from the *multi-resolution* perspective, and then subsequently from the *multi-scale* perspective.

### 2.1. Multi-resolution perspective

#### 2.1.1. Lattice setup

Compared to our previous research [12, 13, 16], the *multi-resolution* concept is an innovative design which is based on two different resolution lattices within the microscopic environment. One is a 100x100 low-resolution lattice with a grid size length of approximately 62.5 μm, reflecting the smallest unit of a hemocytometer used in comparable *in vitro* experimental setups for counting the number of cells via light microscopy (**Figure 1(a)**). The other is a 6x6 high-resolution lattice (superimposed on each of the grid points of the aforementioned low-resolution lattice) with a grid size of approximately 10 μm, thus reflecting the idealized diameter of a cancer cell. We also define that one grid point of the low-resolution lattice can contain only one single cell cluster, and that one cancer cell occupies one grid point of the high-resolution lattice only. This implies that a cell cluster can consist of 36 cancer cells at most, denoted as a *dense* cluster. With the multi-resolution lattice configuration depicted in **Figure 1(b)**, the model can simultaneously simulate the cancer's progression and invasion at two different spatial resolutions: (a) **Low resolution**: This model visualizes, at any point in time, the entire (2D) tumor growth by displaying the behavior of both *active* and *inactive* clusters in the microscopic environment. (b) **High resolution**: In this model, an *active* cluster is considered as a heterogeneous cell cluster in which the cells can possess a migratory, proliferative, or quiescent phenotype, while an inactive cluster is considered as a homogeneous cell cluster in which all the cells are in a quiescent phenotype. For this reason, we only evaluate the molecular network profile in cells that belong to active clusters in an effort to reduce the computational workload and to model the cells' phenotypic fate on the high-resolution lattice. The flowchart of the resulting cellular automaton is pictured in **Figure 2**. Note that **Figure 2 (b)** is used to illustrate step 3 of **Figure 2 (a)** in more detail.

**Figure 1**

**Figure 2**

### 2.1.2 Computational methods

We developed four different computational methods to classify the cells into *inactive* or *active* clusters, and a control to serve as the baseline for comparison.

1. *Control*: All clusters are set to active so that each cell is tracked by the program; that is, each of these cell's molecular signaling pathway is being simulated at every point in time so that the fluctuating concentrations of its (environmentally influenced) sub-cellular signal processing components trigger dynamic phenotypic changes throughout the cells.

2. *Space method*: If all the topographic neighborhoods of a *dense* cluster are themselves *dense*, then this cluster is deemed inactive; otherwise, it is active.

3. *Radius method*: At each time step, first the average distance to the center of mass of all active clusters is set as the basic radius threshold $R_b$. We then also calculate the tumor progression radius difference between the control tumor and the one generated by this method ($DIFF_{RC} = \frac{Radius_{\text{radius method}} - Radius_{\text{control}}}{Radius_{\text{control}}}$) to dynamically adjust the radius threshold with $R_A = R_b \cdot (1 + DIFF_{RC}) \cdot weight$. Thus, at each time step, if a cluster's distance to the center-of-mass is less than $R_A$, that cluster is set to be inactive. Otherwise, it is deemed active. This follows the concept that surface asymmetries must be monitored more closely in a growing tumor as they reflect an intrinsic and/or extrinsic heterogeneity with potential implications for the overall dynamics.

4. *Number method*: At each time step, all clusters are ranked in a descending ordered list according to the distance between each of the clusters and the center of mass of all the clusters. First, half of the clusters are set to the basic number threshold $N_b$. Secondly, we again employ the tumor progression radius difference between the control and number method ( $DIFF_{NC} = \frac{Radius_{number\,method} - Radius_{control}}{Radius_{control}}$ ) to dynamically adjust the threshold with $N_A = N_b \cdot (1 - DIFF_{NC})$. At each time step, if its number is greater than $N_A$, the cluster is set as an inactive cluster. Otherwise, it is considered to be active.

5. *Phenotype method*: At each time step for every cluster, first $P_b$ is set as the basic threshold ratio of the sum of migratory and proliferative cells to the total cells in the cluster. We then employ the tumor progression radius difference between the control and the phenotype method ( $DIFF_{PC} = \frac{Radius_{phenotype\,method} - Radius_{control}}{Radius_{control}}$ ) to dynamically adjust the ratio threshold according to $P_A = P_b \cdot (1 + DIFF_{NC})$. At every time step, if the ratio of the sum of the number of migratory and proliferative cells to the total number of cells in the cluster is less than $P_A$, then that cluster is deemed inactive. Otherwise, it is considered to be active.

**2.2. Multi-scale perspective**

Here, we introduce briefly how the model simulates both tumor progression and invasion on and across multiple scales (for more details, please see [16, 17, 29]). Based on a solid body of work implicating its role in tumor progression at the molecular level, each cell is equipped with a simplified epidermal growth factor receptor or *EGFR* pathway (**Figure 3**). Specifically, **equations 1-19** in **Table 1 (a)** are employed to describe the chemical reactions among the molecular species in this pathway with **Table 1 (b)-(d)** listing the corresponding parameter values [31].

**Figure 3**
**Table 1**

Based on the proposed dichotomy between the migratory and proliferative phenotypes in glioma by Giese et al. [32], and the observation of a transient increase in *PLCγ* resulting in (breast) cancer cell migration by Dittmar et al. [33], we have previously [12, 13, 16] hypothesized the following simplified biological switching behavior: if the percent change of the glioma cell's phosphorylated *PLCγ* concentration exceeds a set change rate threshold of the concentration of phosphorylated *PLCγ*, the cell becomes migratory; otherwise, it adopts a proliferative phenotype. Here, we use the average phosphorylated *PLCγ* concentration changes of all the cells as the change rate threshold at each time step. (We note that *PLCγ* is merely meant as a representative example of a presumably much more complicated set of molecular switching profiles).

On the micro-macroscopic level, we employ a *continuum* module to simulate the diffusion of the chemical cues glucose ($X_0$) and $TGF_\alpha$ ($X_1$) with **equations 20** and **21**:

$$\frac{\partial X_0^{ij}}{\partial t} = D_G \cdot \nabla^2 X_0^{ij}, t = 1,2,3... \qquad (20)$$

$$\frac{\partial X_1^{ij}}{\partial t} = D_T \cdot \nabla^2 X_1^{ij}, t = 1,2,3... \qquad (21)$$

where $D_G$ is the diffusion coefficient of glucose [34], and $D_T$ is the $TGF_\alpha$ diffusion coefficient [35]. Furthermore, we employ a *discrete* module to simulate the cell's glucose uptake and $TGF_\alpha$ secretion on the high-resolution lattice with **equations 22** and **23**:

$$X_0^t = X_0^{t-1} - U_G \qquad (22)$$

$$X_1^t = X_1^{t-1} + S_T \qquad (23)$$

where *t* represents the time step, $U_G$ is the cell's glucose uptake coefficient [36], and $S_T$ is the $TGF_\alpha$ secretion rate [37]. The diffusions of glucose and $TGF_\alpha$ are simulated on the low-resolution lattice; however, the concentration of $TGF_\alpha$ and glucose at each grid point in the low-resolution lattice will be randomly distributed on the corresponding high-resolution lattice.

## 3. RESULTS

Our code is written in Microsoft Visual Studio C++. We ran the simulation 10 times with varying migration durations (1-5 time steps, one time step being equivalent to one hour) for each computational method. The algorithm requires approximately 1 hour of CPU time on a Dell Precision 690 workstation with 64-bit Quad-Core Intel® Xeon® 5300 series processors.

**Microscopic patterns. Figure 4** displays the snapshots of the virtual tumor cells at time steps 1 and 100 for the control **(a)** and the four computational methods **(b-e)**, respectively. Here it should be noted that since different computational methods employ varying classification schemes to separate the cancer cells into either active or inactive clusters, **Figure 4** shows qualitative differences among these methods at the first time step. At the start, for example, the control and phenotype method will classify all the cells into active clusters, whereas the number, radius, and space methods will separate the cells into active and inactive clusters.

**Figure 4**

**Performance. Figure 5 (a)** displays the computation time required for each multi-resolution method, with the control serving as a comparison. While the number method, as already qualitatively suggested by **Figure 4**, indeed requires the minimum computation time, unfortunately its resulting tumor radius progression showed the largest deviation from the control (**Figure 5 (b)**), fluctuating considerably and exhibits a substantial average error (**Figure 5 (c)**). Conversely, while the space, radius, and phenotype methods also resulted in reduced computation time (**Figure 5 (b)**), their tumor radii remained very close to the control's (and rather stable throughout), thus yielding a relatively small average error (**Figure 5 (c)**).

**Figure 5**

**Functionality. Figure 6 (a)** and **(b)** describe the number of active and inactive clusters for each computational method. As intended, the control only operates with active clusters and thus represents the upper limit. Also, **Figure 4**'s qualitative impression is now quantitatively supported, in that the number method indeed utilizes the lowest number of active clusters, and the highest number of inactive clusters. The space and radius methods operate with moderate numbers of active and inactive clusters. With regards to the phenotype method, we found its profile of active and inactive clusters to evolve in a similar fashion to the control during the early stages up until time step 50, after which its number of active clusters drops abruptly, and the number of inactive clusters increases markedly, subsequently followed by a period of stability. Likewise, **Figure 6 (c)** and **(d)** show the number of checked and unchecked cells, which parallel the findings on the cluster level.

**Figure 6**

**Ranking.** To integrate the findings presented in **Figures 4-6** in a quantitative manner, we have developed a formula that ranks the overall performance of each computational method according to

$$R = \sum_{i=1}^{n} \frac{DIFF_i * Time_j}{n * Time_{control}}, i = 1,2,3,\cdots 100, j = 1,2,3,4, n = 100. \qquad (24)$$

where *i* is the time step from 1 to 100, and *j* represents the specific computational method with (1) being the space method, (2) the radius method, (3) the number method, and (4) the phenotype method. *Inactive*$_i$ and *active*$_i$ are the numbers of inactive and active clusters, respectively. $DIFF_i$ is the simulation difference in tumor radius between each computational method and the control, represented by $DIFF_i = \frac{Radius_i - Radius_{control}}{Radius_{control}}$. Here, the best performing method should use *less* computation time by operating with a *greater* number of inactive clusters and a *lower* number of active clusters, while simultaneously only having to accept a small difference in the resulting tumor progression radius as compared to the control. As listed in **Table 2**, using this method, the radius method ranks as number one, followed by the space and phenotype methods; the number method ranks last.

**Table 2**

### 4. DISCUSSION & CONCLUSIONS

Mathematical modeling and computational simulations have become increasingly important for data integration as well as hypothesis generation in cancer research. Methods used include a variety of techniques. For example, Cristini et al. [38] employed a *continuum* mathematical approach to describe microenvironmental substrate gradients that may drive morphologic instability in gliomas, while Jain et al. [39] used the same modeling technique to simulate angiogenesis under the control of the *VEGF–Bcl-2–CXCL8* pathway. Conversely, Peirce et al. [40] employed a *discrete* method (cellular automaton [25]) to predict micro-vascular network patterning by integrating epigenetic stimuli, molecular signals, and cellular behaviors, and Stamatakos et al. [11], also using a discrete approach, simulated the glioma cell cycle explicitly to predict the response of radiotherapy. In contrast to both approaches, Mansury et al. [14, 15], Athale et al. [12, 13] and Zhang et al. [16, 29] all employed a discrete-continuum (i.e. *hybrid*) model to simulate brain cancer progression. From these examples, in addition to other works, it still holds that while continuum modeling can appropriately represent tissue level properties [38, 41] and cell densities [4, 20, 42, 43], it is impractical to describe single entities such as cells, genes, or proteins and their various interactions on the cellular and sub-cellular level. However, to achieve such resolution with a discrete model requires significant computational resources particularly when simulating millions of cells simultaneously. And, while hybrid modeling emerges as the most promising avenue in that it is capable of simulating both tumor tissue and single cell behavior at the same time, it mandates a reduced description of microenvironmental complexity and is bound by the number of simulated cells and signaling pathways to remain computationally tractable.

To begin to address the currently limited *scalability* of such hybrid computational models, we have developed a *multi-scale, multi-resolution agent-based* glioma model that classifies the simulated cancer cells into either 'active' or 'inactive' clusters with several novel computational methods, such that it is possible to utilize currently limited computational capabilities to simultaneously model millions of cells. As mentioned before in the result section, these computational methods are designed with respect to micro-macroscopic properties such as peritumoral space, tumor radius, and tumor cluster number, as well as cellular properties such as the phenotypes' sub-cellular signaling networks trigger. The major aim of this research has been to demonstrate that these computational methods can reduce the computational workload to a certain extent while preserving predictive power. Additionally, the simulation results of these computational methods show several interesting findings. First, the space and radius methods result in tumor expansion patterns that are qualitatively similar to the control, more so than the phenotype and number methods (**Figure 4, Figure 5 (b)** and **(c)**). Second, comparing the space, radius, number and phenotype methods versus control, we found a 9.34%, 1.95%, 36.74% and 7.74% reduction of the necessary computation time, respectively. Thus, the number method commands the least computation time (**Figure 5 (a)**), generates fewest active clusters, and checks the fewest cells as compared to the other methods, including the control (**Figure 6**). Third, the training algorithms are superimposed on these computational methods except the space method to enhance the predictive power, but the tumor radii from every method is still not able to exactly match that of the control (**Figure 5b**, **Figure 7c**). While the resulting "average error vs. control" (**Figure 5c**, **Figure 7d**) for most of the computational methods remain relatively small throughout the length of the simulation, it is especially apparent with the number method, exhibiting the weakest predictive power as measured by the substantial "average error vs. control" value. Lastly, since there is no optimal computational method here that simultaneously yields the least radial deviation from the control while *also* uses the least amount of computation time, we have developed a novel formula (**equation 24**) to integrate considerations from the vantage points of both predictive power and computational workload. This equation was then used to rank the efficiencies of all the computational methods, with **Table 2** showing the following result: the radius method tops the performance scale, followed by the space and phenotype methods, and lastly the number method.

This ranking result can be explained as follows: For the cellular automaton component of our model [15, 16] (**Figure 2**), a cancer cell will choose an unoccupied lattice site with the highest glucose concentration to divide into or move to. Once there are no unoccupied locations in its neighborhood, the cell will turn quiescent. Therefore, both the space and radius method save computation time by deactivating the tightly packed cells in the center of the tumor. In doing so, the radius method is superior to the space method, because deactivating the clusters in the boundary of the tumor (space method) will negatively impact the tumor progression radius more so than deactivating the clusters in the center of the tumor (radius method). However, the classification mechanism of the space and radius methods is different from that of the number method. That is, because the shape of the tumor is irregular and the density of the clusters is inhomogeneous, the number method will deactivate many cells within the vicinity of the tumor boundary, weakening the result by way of a large "average error vs. control" (**Figure 5 (c)**, **Figure 7 (d)**). With regards to the phenotype method, especially at the beginning of the simulation, the algorithm must check each cell's molecular pathway to decide if it is an active or an inactive cluster (similar to the process in the control). The cell cycle requires several time steps to switch a cell's phenotype, which (in **Figure 6**) leads to the marked changes around time step 50 and hampers the phenotype method's capacity to reduce the computational workload in the early stages of the simulation. Because there is no ideal algorithm, amongst the ones studied, that combines predictive power with low computation time, we presented a new method by integrating the #1 and #2 ranked methods (**Table 2**) (i.e. adding the radius requirement into the space method; the other way around would not improve upon the radius method's singular performance, due to the underlying criteria) in an effort to determine if such a combinatorial method would further improve the results, and to demonstrate the flexibility and extensibility of this study.

For the combination method, we use the radius method to first classify the cells in the center of the tumor into inactive clusters, after which only the clusters in the boundary of the tumor are left to be classified by the space method. We hypothesize that this approach reduces computation time markedly. And indeed, **Figure 7 (a)** confirms that the computation time of this combination method exhibits a 14.1% reduction versus the control, substantially shorter than both the space and radius methods *alone* due to the generation of a lower number of active (and a greater number of inactive) clusters (**Figure 7 (b)** and **Figure 6**). However, **Figure (c)** shows also that the error (vs. control) generated by the combination method exceeds that resulting from the radius method, but remaining lower than that of the space method. This is due to the fact that although employing both radius and space requirements can save computation time, the clusters on the boundary of the tumor that are deactivated by the space method will still influence the tumor progression curve. Taken together, however, our ranking system establishes this combination method as the new optimal performing method, even if its rank value of 4.06774 is only a relatively small improvement over the radius method alone.

**Figure 7**

Despite its technical merits, the current approach still has several pitfalls. At this point, although the training algorithms of these computational methods can adjust the active/inactive cluster number threshold at each time step by the "average error vs. control" (**Figure 5 (c)**) in the previous time step (Section 2.1.2 illustrates this in detail), the cell cycle duration delays cell proliferation in the *active* clusters such that these active clusters cannot have sufficient migratory or proliferatory cells in the following step. The result is that the tumor radius never really approaches that of the control (**Figure 5 (b)**), with this error never really becoming smaller (**Figure 5 (c)**). To address these shortcomings in the future, we anticipate that the algorithms will have to be trained on other attributes such as topographic surface patterns, the tumor cell number, and perhaps signaling pathway profiles that are implicitly related to the cell cycle, alongside the tumor progression radius. This way, the impact of the time delay may be reduced.

In summary, our previously developed 2D brain tumor model has been given a novel multi-resolution design. This allows the incorporation of several computational methods, and thus their performance testing, in an effort to reduce the computational workload while still maintaining sufficiently high predictive power. The multi-resolution design provides this 2D model, which only incorporate roughly 3000 cells over relatively small ECM volumes and short simulation durations, with a small advantage. However, a future goal is the prediction of actual tumor progression, with a model consisting of millions of cells over much larger ECM volumes and longer simulation durations. It is thus reasonable to expect that applying these methods to a more clinically-relevant 3D model will save even more computational resources over time, which would help in bringing these promising computational approaches ever closer to the clinical arena.

## ACKNOWLEDGEMENTS

This work has been supported in part by NIH grants CA 085139 and CA 113004 (The Center for the Development of a Virtual Tumor, CViT at https://www.cvit.org) and by the Harvard-MIT (HST) Athinoula A. Martinos Center for Biomedical Imaging and the Department of Radiology at Massachusetts General Hospital.

# CAPTIONS

**Figure 1: (a)** Hemocytometer used in a typical *in vitro* cell counting experiment, with the minimum length of the grid size being 62.5 μm. **(b)** A schematic of the coupled high- and low-resolution lattices.

**Figure 2:** Algorithm flow charts **(a)** for classifying the cells into either active or inactive clusters on the low-resolution lattice, and **(b)** for checking each cell's molecular pathway in an active cluster on the high resolution lattice.

**Figure 3:** The EGFR gene-protein interaction network [17, 31].

**Figure 4:** Snapshots of the simulation for each computational method at time step 1 (*left*) and 100 (*right column*): **(a)** control, **(b)** space method, **(c)** radius method, **(d)** number method, and **(e)** phenotype method. Note that *black* represents active clusters (comprised of proliferative, migratory and quiescent cells), and *green* represents inactive clusters.

**Figure 5:** Shown are **(a)** the average computation time of 10 runs with bars representing standard deviation, and **(b)** the simulated tumor radius for the four multi-resolution methods and the high-resolution control. The *x-axis* denotes the time step, and the *y-axis* represents the tumor radius (in pixels; 1 pixel = 1.47μ.) **(c)** Average error of the space method, radius method, number method, and the phenotype method versus control. The *x-axis* denotes the time step, and the *y-axis* represents the percent deviation for a particular method from the control.

**Figure 6:** Depicts **(a)** the number of active multi-cellular clusters, **(b)** the number of inactive multi-cellular clusters, **(c)** the number of checked cells and **(d)** the number of unchecked cells. (Note: If the program investigates a cell's molecular pathway, this cell is deemed 'checked', otherwise it is regarded as 'unchecked').

**Figure 7:** Shown are **(a)** the average computation times of 10 runs with bars representing the standard deviation for the "combination" method, **(b)** the number of active and inactive multi-cellular clusters for this method versus control, **(c)** the simulated tumor radius for the five multi-resolution methods including the combination method versus high-resolution control.

Here again, the *x-axis* denotes the time step, and the *y-axis* represents the tumor radius (in pixels; 1 pixel = 1.47μm), and **(d)** the average error between the space, radius, number, phenotype, or combination methods and the control. The *x-axis* denotes the time step, and the *y-axis* represents the percent deviation for a particular method from the control.

**Table 1: (a)** Kinetic equations employed to describe the reactions between the *EGFR* species; **(b)** components of the *EGFR* gene-protein interaction network; **(c)** coefficients of the *EGFR* gene-protein interaction network taken from the literature [17, 31]; **(d)** coefficients of the micro-macroscopic environment of the model [16, 17, 28, 36].

**Table 2:** The rankings for **(1)** the radius method, **(2)** the space method, **(3)** the phenotype method, and **(4)** the number method, using **equation 24**.

# FIGURES & TABLES

**Figure 1**

**(a)**

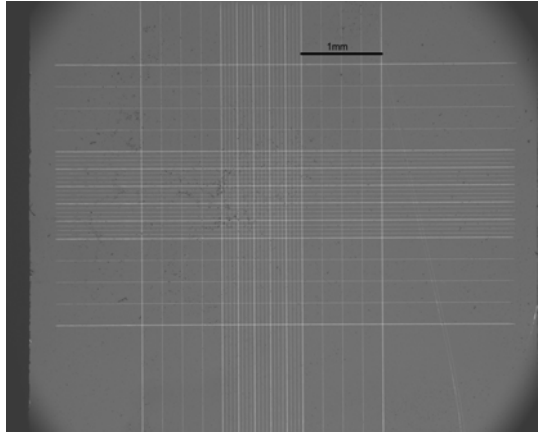

**(b)**

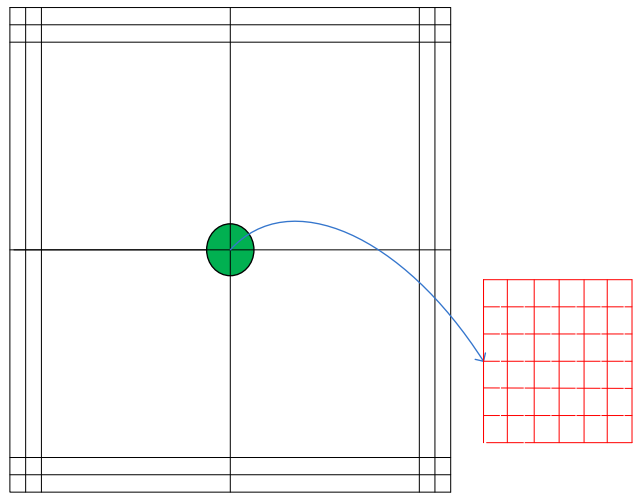

**Figure 1**

**Figure 2**

**(a)**

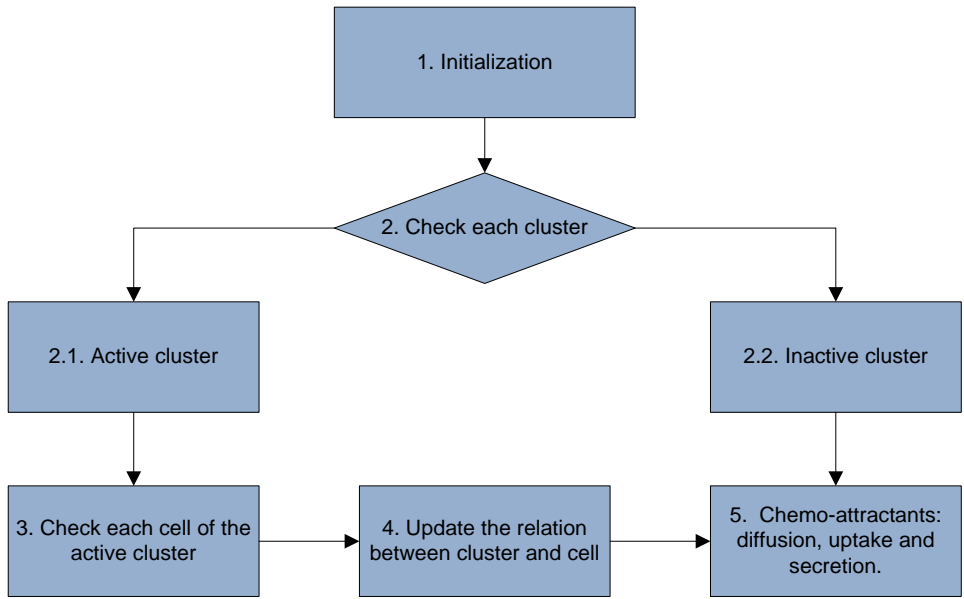

**(b)**

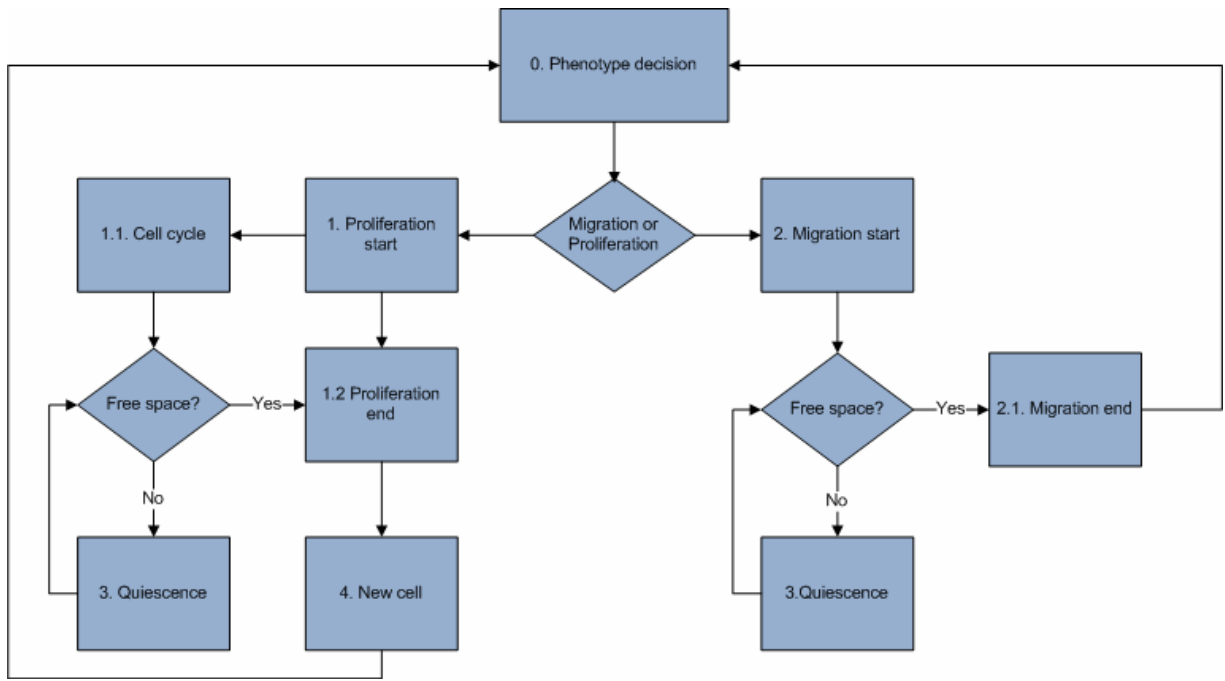

**Figure 3**

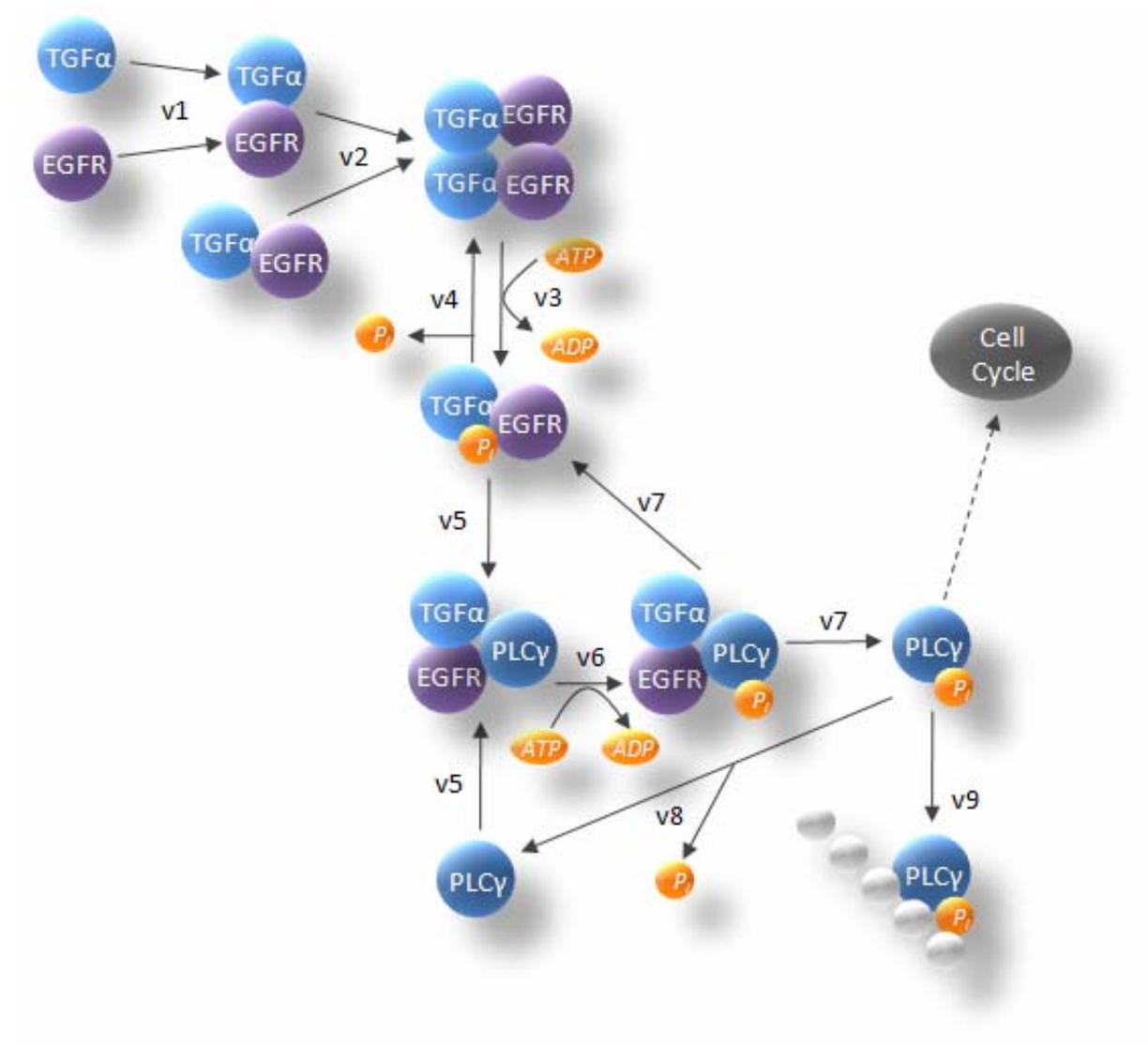

**Figure 4**
**(a)**
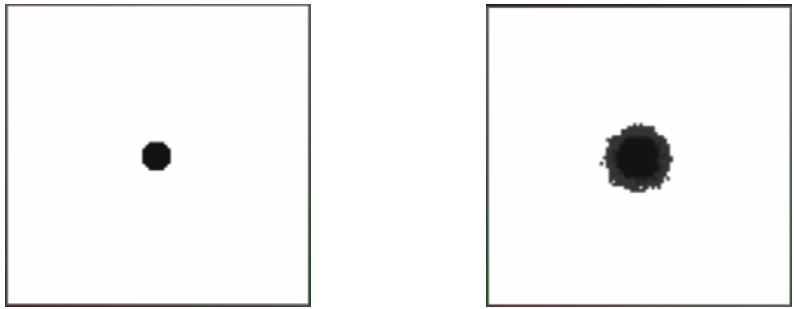
**(b)**
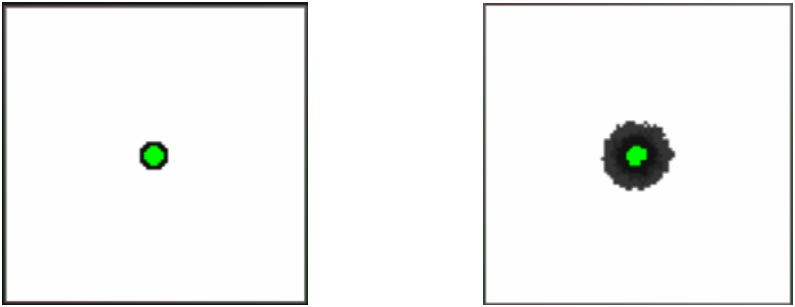
**(c)**
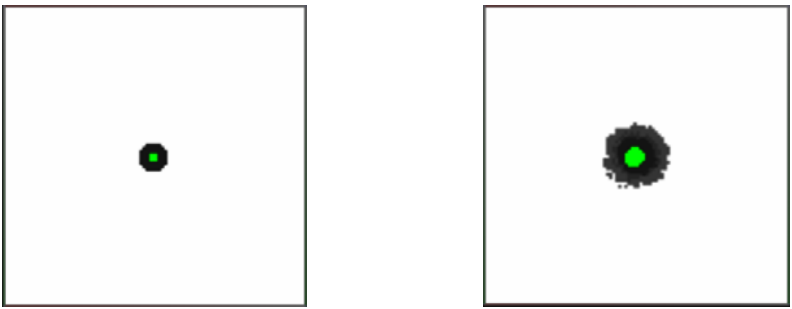
**(d)**
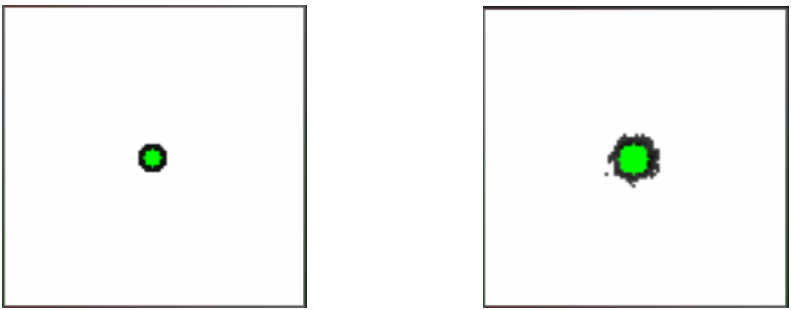
**(e)**
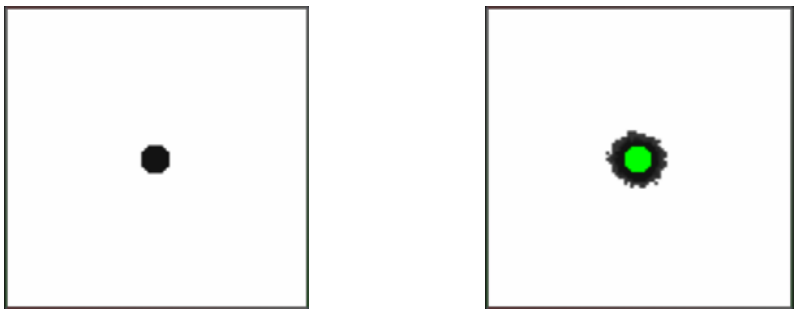

**Figure 5**

**(a)**

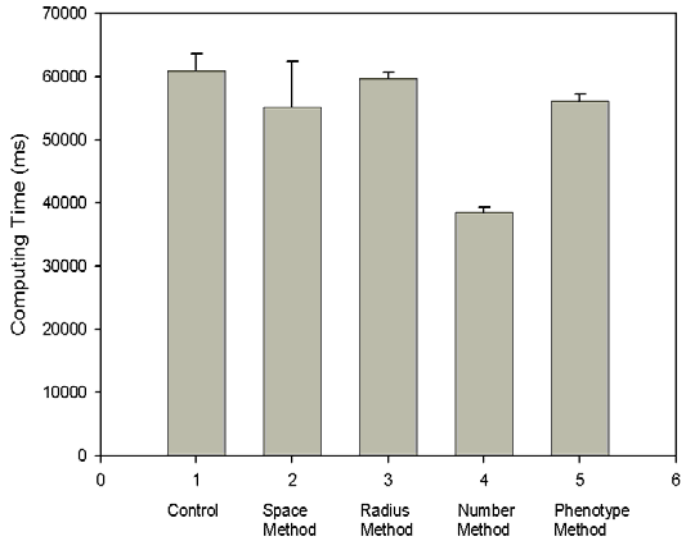

**(b)**

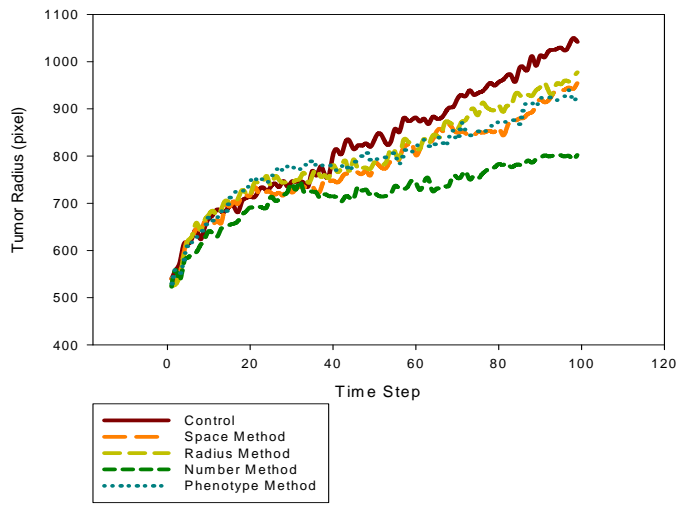

**(c)**

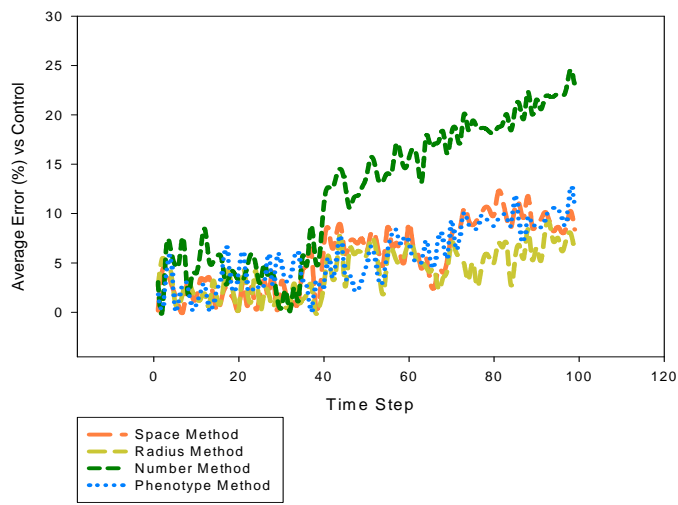

**Figure 6**

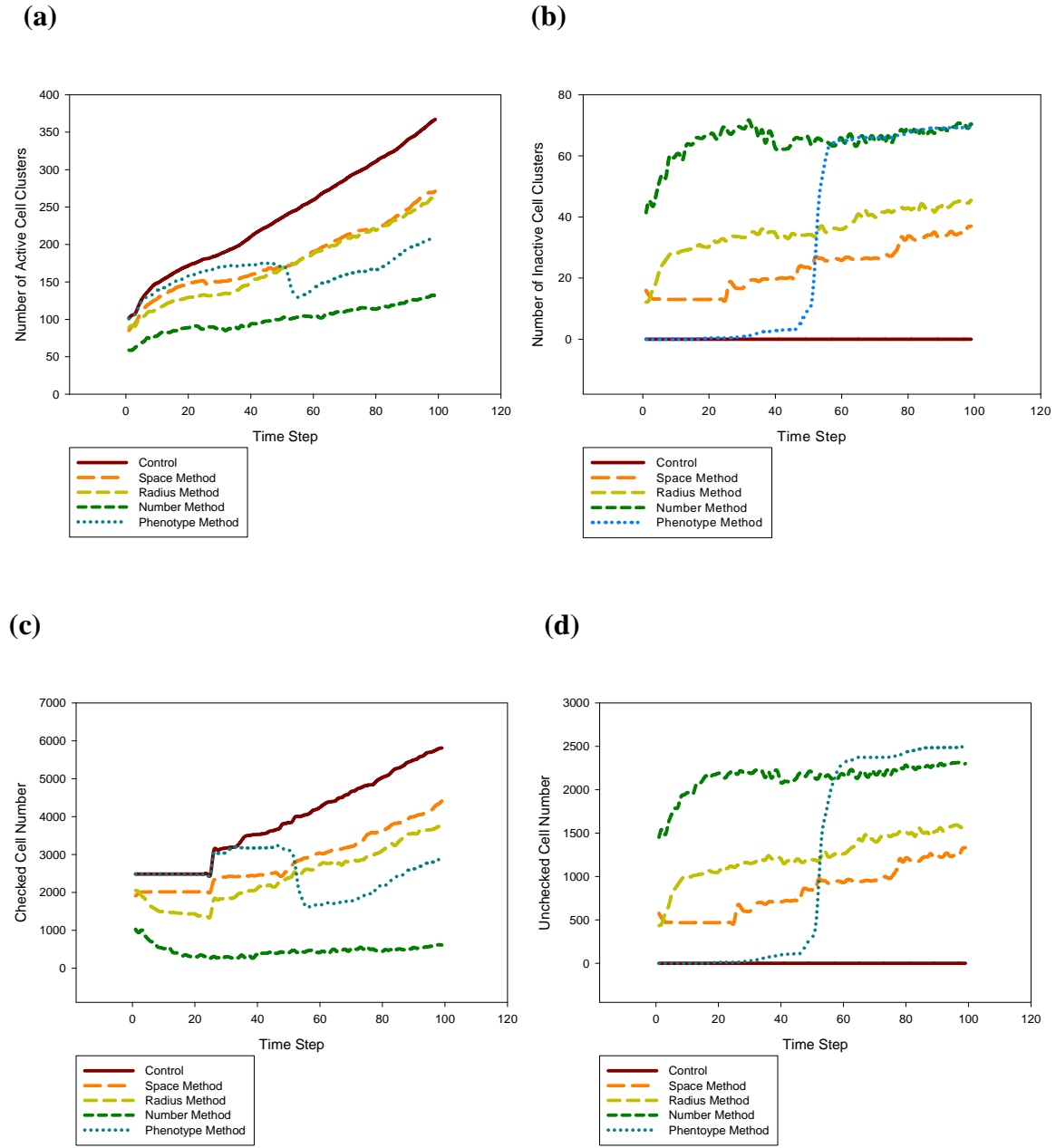

**Figure 7**

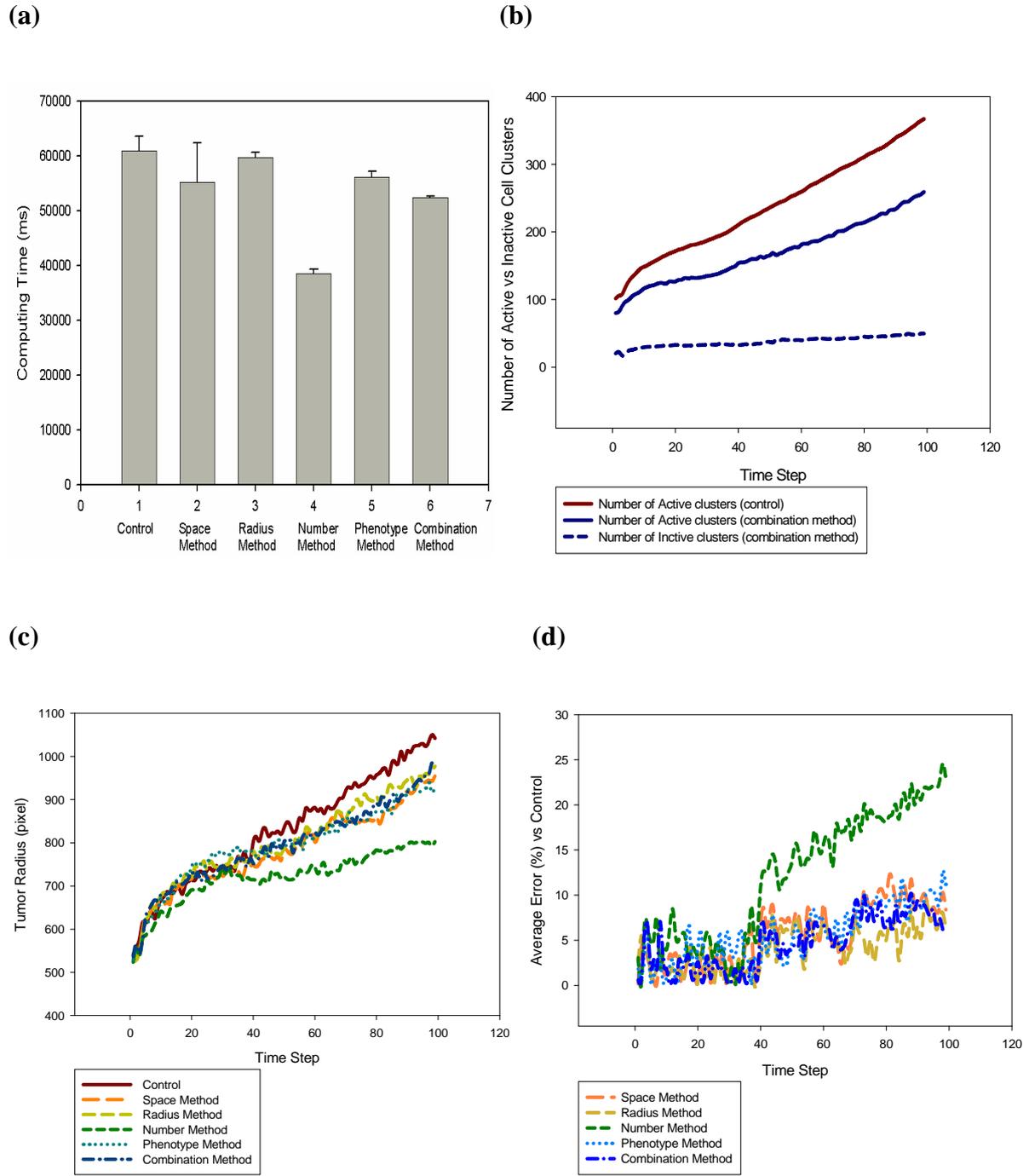

# TABLES

**Table 1**

**(a)**

| | | | |
|---|---|---|---|
| $dX_1/dt = -v_1$ | (1) | $v_1 = k_1 X_1 X_2 - k_{-1} X_3$ | (11) |
| $dX_2/dt = -v_1$ | (2) | $v_2 = k_2 X_3 X_3 - k_{-2} X_4$ | (12) |
| $dX_3/dt = v_1 - 2v_2$ | (3) | $v_3 = k_3 X_4 - k_{-3} X_5$ | (13) |
| $dX_4/dt = v_2 + v_4 - v_3$ | (4) | $v_4 = V_4 X_5 / (K_4 + X_5)$ | (14) |
| $dX_5/dt = v_3 + v_7 - v_4 - v_5$ | (5) | $v_5 = k_5 X_5 X_6 - k_{-5} X_7$ | (15) |
| $dX_6/dt = v_8 - v_5$ | (6) | $v_6 = k_6 X_7 - k_{-6} X_8$ | (16) |
| $dX_7/dt = v_5 - v_6$ | (7) | $v_7 = k_7 X_8 - k_{-7} X_5 X_9$ | (17) |
| $dX_8/dt = v_6 - v_7$ | (8) | $v_8 = V_8 X_9 / (K_8 + X_9)$ | (18) |
| $dX_9/dt = v_7 - v_8 - v_9$ | (9) | $v_9 = k_9 X_9 - k_{-9} X_{10}$ | (19) |
| $dX_{10}/dt = v_9$ | (10) | | |

**(b)**

| Symbol | Molecular variables | Initial values |
|---|---|---|
| $X_0$ | *Glucose* | 25 *mM* |
| $X_1$ | *TGFα* | 9010.55 *nM* |
| $X_2$ | *EGFR* | 100 *nM* |
| $X_3$ | *TGFα –EGFR* | 0 *nM* |
| $X_4$ | *(TGFα –EGFR) 2* | 0 *nM* |
| $X_5$ | *TGFα –EGFR-P* | 0 *nM* |
| $X_6$ | *PLCγ* | 10 *nM* |
| $X_7$ | *TGFα –EGFR-PLCγ* | 0 *nM* |
| $X_8$ | *TGFα –EGFR-PLCγ-P* | 0 *nM* |
| $X_9$ | *PLCγ-P* | 0 *nM* |
| $X_{10}$ | *PLCγ-P-I* | 0 *nM* |

**(c)**

| Forward rate ($s^{-1}$) | Reverse rate ($s^{-1}$) | Michaelis constants ($nM$) | Maximal enzyme rates ($nM\ s^{-1}$) |
|---|---|---|---|
| $k_1 = 0.003$ | $k_{-1} = 0.06$ | $K_4 = 50$ | $V_4 = 450$ |
| $k_2 = 0.01$ | $k_{-2} = 0.1$ | $K_8 = 100$ | $V_8 = 1$ |
| $k_3 = 1$ | $k_{-3} = 0.01$ | | |
| $k_5 = 0.06$ | $k_{-5} = 0.2$ | | |
| $k_6 = 1$ | $k_{-6} = 0.05$ | | |
| $k_7 = 0.3$ | $k_{-7} = 0.006$ | | |
| $k_9 = 1$ | $k_{-9} = 0.03$ | | |

**(d)**

| Coefficient | Value | Units | Description |
|---|---|---|---|
| $D_G$ | $6.7 \times 10^{-7}$ | $cm^2 s^{-1}$ | Glucose diffusion coefficient |
| $D_T$ | $5.18 \times 10^{-7}$ | $cm^2 s^{-1}$ | TGFα diffusion coefficient |
| $S_T$ | $0.2$ | $nMh^{-1}$ | TGFα secretion rate |
| $U_G$ | $7.7 \times 10^{-12}$ | $mol/h.cell$ | Glucose uptake rate |

**Table 2**

| Method | Rank | Ranking value |
|---|---|---|
| Radius | 1 | 4.068785 |
| Space | 2 | 5.139941 |
| Phenotype | 3 | 5.223518 |
| Number | 4 | 7.710731 |